\documentclass[conference]{IEEEtran}
\IEEEoverridecommandlockouts
\pdfoutput=1
\usepackage{cite}
\usepackage{amsmath,amssymb,amsfonts}
\usepackage{algorithmic}
\usepackage{graphicx}
\usepackage{textcomp}
\usepackage{xcolor}
\usepackage{booktabs,etoolbox}
\usepackage{epsfig,subfigure}
\usepackage{amssymb, amsmath}
\usepackage{color}
\usepackage{mathrsfs}
\usepackage{graphicx}
\usepackage{url}

\usepackage{amssymb}
\usepackage{amsmath,amsfonts,amssymb}
\usepackage{verbatim}
\usepackage[bookmarks=false]{}
\usepackage[font={small}]{caption}
\def\BibTeX{{\rm B\kern-.05em{\sc i\kern-.025em b}\kern-.08em
    T\kern-.1667em\lower.7ex\hbox{E}\kern-.125emX}}
    
 \newcommand{\T}[0]{^{\mathrm{T}}}

\begin{document}

\title{Experimental Study on Probabilistic ToA and AoA Joint Localization in Real Indoor Environments
\thanks{C. Geng was with Nokia Bell Labs when he finished his contribution to this work. \\
© 2021 IEEE.  Personal use of this material is permitted.  Permission from IEEE must be obtained for all other uses, in any current or future media, including reprinting/republishing this material for advertising or promotional purposes, creating new collective works, for resale or redistribution to servers or lists, or reuse of any copyrighted component of this work in other works.}
}

\author{\IEEEauthorblockN{Chunhua Geng}
\IEEEauthorblockA{\textit{MediaTek USA Inc.} \\
Irvine, CA, USA \\
chunhua.geng@mediatek.com}
\and
\IEEEauthorblockN{Traian E. Abrudan}
\IEEEauthorblockA{\textit{Nokia Bell Labs} \\
Espoo, Finland \\
traian.abrudan@nokia-bell-labs.com}
\and
\IEEEauthorblockN{Veli-Matti Kolmonen}
\IEEEauthorblockA{\textit{Nokia Bell Labs} \\
Espoo, Finland \\
veli-matti.kolmonen@nokia-bell-labs.com}
\and
\IEEEauthorblockN{Howard Huang}
\IEEEauthorblockA{\textit{Nokia Bell Labs} \\
Murray Hill, NJ, USA \\
howard.huang@nokia-bell-labs.com}
}

\maketitle

\begin{abstract}
In this paper, we study probabilistic time-of-arrival (ToA) and angle-of-arrival (AoA) joint localization in real indoor environments. To mitigate the effects of multipath propagation, the joint localization algorithm incorporates into the likelihood function Gaussian mixture models (GMM) and the Von Mises-Fisher distribution to model time bias errors and angular uncertainty, respectively.  We evaluate the algorithm performance using a proprietary prototype deployed in an indoor factory environment with infrastructure receivers in each of the four corners at the ceiling of a 10 meter by 20 meter section. The field test results show that our joint probabilistic localization algorithm significantly outperforms baselines using only ToA or AoA measurements and achieves 2-D sub-meter accuracy at the 90$\%$-ile. We also numerically demonstrate that the joint localization algorithm is more robust to synchronization errors than the baseline using ToA measurements only.   
\end{abstract}


\section{Introduction}

With the proliferation of ubiquitous wireless devices, ranging from sensors to cell phones to VR/AR equipment to robots, the capability of determining the device positions in complex indoor environments has becomes  integral in modern wireless networks. Indoor localization enables wide-scale applications and services, including indoor navigation, warehouse asset tracking and management, contextual-aware marketing and customer assistant, building surveillance, location based health services, among others \cite{IndoorLocTut}. For this reason, it has attracted considerable research interest from both academia and industry in the past decade.

One of the fundamental challenges in wireless indoor localization is multipath propagation. Due to reflections and diffraction by walls and conductive objects in the indoor environments, multiple replicas of the same transmitted wireless signal may arrive at receivers with different delays and complex gains, and from multiple angles w.r.t. line-of-sight (LOS). Consequently, the harsh propagation conditions pose serious challenges in deconvolving the LOS component, and leads to significant consequences for localization performance. For instance, in the widely-accessible time-of-arrival (ToA) localization systems \cite{GNSS_Book, Cellular_Loc_Survey, UWB_ranging}, multipath introduces positive channel biases, which degrade the localization accuracy significantly in hostile environments. Similarly, for angular estimation in angle-of-arrivial (AoA) localization, multipath appears as a combination of several coherent signals arriving at the antenna array at different angles, which makes the angular estimation very challenging. 
 
To effectively mitigate channel bias errors, a Bayesian probabilistic algorithm has been introduced recently \cite{Blade} for ToA localization, where the channel bias is modeled as a random variable (RV) following Gaussian mixture models (GMM), and incorporated into a maximum-a-posterior (MAP) estimation to determine the device position in a robust way. This algorithm has been generalized from various perspectives. For instance, in \cite{ToA-Hybrid, H-Blade} the probabilistic algorithm has been applied to hybrid positioning with both cellular networks and global navigation satellite systems (GNSS); in \cite{ToA-MVGMM}, it has been extended to account for the channel correlations among different locators; and in \cite{Blade_EP} a computational efficient approach based on expectation propagation \cite{EP} is proposed to solve the non-linear and non-convex MAP estimation.  

In this paper, to improve the indoor localization accuracy we advocate a joint ToA and AoA probabilistic 3-D localization algorithm, and evaluate its performance with a carefully designed prototype system in real indoor environments. In the joint localization algorithm, we leverage the probabilistic approach for ToA positioning in \cite{Blade}, and {\em directional statistics}~\cite{2009_MarJup} to model the uncertainty of the AoA estimates. The adoption of directional statistics here is motivated by the fact that angles are periodic in their nature, i.e., they are defined on a circle of sphere, rather than Euclidean space. Therefore, the standard Gaussian distribution is not the most appropriate. To the best of our knowledge, directional statistics have been used for the first time for 2-D AoA positioning in~\cite{2012_WanJacInk,ToA_AoA_Visa} by employing the von Mises distribution, and for 3-D positioning in~\cite{2016_AbrXiaMarTri,2017_NurSuoPic} by employing the more general von Mises-Fisher distribution.  Other 2-D probabilistic AoA positioning approaches can be found in~\cite{2013_XioJam,2017_BniErgSubSteWey}. 

Notably, in previous studies (e.g., \cite{TDoA_AoA_WCDMA, ToA-AoA-Eu}), joint ToA and AoA localization has been mostly analyzed from a theoretical perspective and evaluated with simulated data in 2-D layout. Specifically,  for the \emph{probability} approach, an example can be found in a very recent study in \cite{ToA_AoA_Visa}. One of the main contributions in our paper is that we experimentally demonstrate the superiority of \emph{probabilistic} joint ToA and AoA localization in real world.  Towards this end, we set up proprietary ToA and AoA localization systems (both having multiple locators) in a real indoor factory environment and assess the joint ToA and AoA localization performance with over-the-air measurements, in order to capture the effects of  real propagation conditions of the complex indoor environments, as well as the hardware imperfections. Real-world measurements have also been used in several previous studies on indoor localization. For instance, in \cite{ 2016_VasKumKat, 2018_SolKalAviWhi, 2019_ReaAbrGiuClaKol, SB_ToA_AoA_LTE}, the localization algorithms using a single locator (i.e., access point in WiFi or base station in LTE) have been evaluated in indoor environments. Specifically, Chronos in \cite{2016_VasKumKat} is based on trilateration to estimate the target position; MonoLoco in \cite{2018_SolKalAviWhi} is based on triangulation with the help of multipath reflections; SPRING in \cite{2019_ReaAbrGiuClaKol} and the LTE testbed in \cite{SB_ToA_AoA_LTE} use angular and ranging measurements to directly compute the target location. In \cite{2015_KotJosBhaKat}, a localization system, named SpotFi, is developed based on jointly processing ToA, AoA, and received signal strength indicator (RSSI) from multiple locators, where the ranging measurements are specifically utilized to help identify the AoA LOS component. It is noteworthy that many aforementioned algorithms (e.g., in Chronos, MonoLoco and SpotFi) rely on  low-level measurements, such as channel state information (CSI) per subcarrier per antenna. In our algorithm, what we need is only high-level ToA and AoA estimations for each locator, which is an advantage since in many applications and services the low-level information like CSI is not disclosed by the vendors. The field test results demonstrate that our joint localization algorithm significantly outperforms the baselines using either ToA or AoA data only, achieving sub-meter level accuracy at 90$\%$-ile for horizontal localization (given inter-locator distances no less than 10 meters). In addition, we also numerically illustrate that compared with the ToA baseline, the joint localization algorithm is much more robust to synchronization errors. 
 

\section{Problem Formulation \label{sec_model}}
Consider a joint ToA and AoA localization system with $K$ ToA locators and $B$ AoA locators. We denote the unknown user device (UD) location by $\mathbf{x}=[x,y,z]$.\footnote{We assume the localization system is an uplink system, where the UD broadcasts the positioning reference signals, and the locators receive the signals and estimate the user's position. The algorithm presented in this paper can be easily adapted for downlink systems.} 
All ToA locators are time-synchronized with each other, but not with the UD. Denote the ToA of the positioning reference signal at $k$-th ToA locator by\footnote{For notation brevity, we convert time to distance by multiplying with the speed of light implicitly.}
\begin{align}\label{rstd_model}
t_{k}=||\mathbf{p}_{k}-\mathbf{x}||+\tau+\gamma_k+n_k,~~\forall k\in\{1,2,\ldots,K\}
\end{align}
where $\mathbf{p}_{k}=[x_{k}, y_{k}, z_{k}]$ is the position of the $k$-th ToA locator (in the World frame), $\tau$ is the unknown transmit time of the reference signal with respect to the clock at locators, $\gamma_k$ represents the channel bias introduced by unresolvable multipath and NLOS reflections, and $n_k\sim \mathcal{N}(0,\sigma^2)$ accounts for both the locator synchronization errors and the ToA measurement error due to thermal noise. Following \cite{Blade}, we assume that the channel bias $\gamma_k$ is a RV following GMM with $L_k$ components, i.e.,
\begin{align}\label{SV-GMM}
p(\gamma_k) = \sum_{i=1}^{L_k}\frac{w_{ik}}{\tilde{\sigma}_{ik}\sqrt{2\pi}}\exp\left[-\frac{1}{2\tilde{\sigma}^2_{ik}}(\gamma_k-\mu_{ik})^2\right]
\end{align}
where $w_{ik}$, $\tilde{\sigma}^2_{ik}$, and $\mu_{ik}$ represent the weight, variance, and mean value of the $i$-th Gaussian component for the $k$-th locator.

\begin{figure}
    \centering
    \includegraphics[width= 6.6 cm]{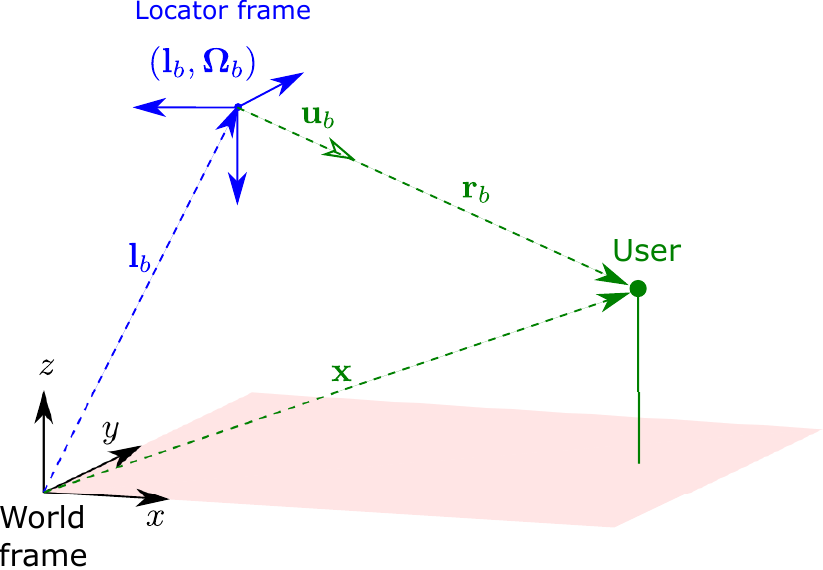}
    \caption{Illustration of the 3-D angular positioning.}
    \label{fig:aoa_model}
\end{figure}

The model for angular positioning is illustrated in Fig.~\ref{fig:aoa_model}. We consider $B$ AoA locators whose 3-D positions are represented in the World frame by a $3 \times 1$ vectors ${\mathbf{l}}_b$, and whose 3-D orientations are represented by a $3 \times 3$ orthogonal matrices ${\boldsymbol{\Omega}}_b$. Both the positions and orientations of all $B$ AoA locators ($b\in\{1,2,\ldots,B\}$) are assumed to be known with reasonable accuracy. The AoA locators are equipped with phased antenna arrays that are able to estimate the directions of the incoming signals. These directions are defined w.r.t the local frame of coordinates of each of the locators. Let us denote the user's position vector in the locator's frame by ${\mathbf r}_b$. The user's position vector expressed in the World frame can be  written as 
\begin{equation} \label{aoa_model_equation}
{\mathbf x} = {\mathbf l}_b + {\boldsymbol{\Omega}}_b {\mathbf r}_b
\end{equation}
Since there is no range information available at the AoA locator, the length of ${\mathbf r}_b$ is unknown, i.e., only its direction ${\mathbf u}_b={\mathbf r}_b/\|{\mathbf r}_b\|$ can be estimated.
From Eq.~\eqref{aoa_model_equation}, we can express the true direction of the user as a function of user's true position ${\mathbf x}$, as follows
\begin{equation} \label{true_direction}
{\mathbf u}_b({\mathbf x}) = {\boldsymbol{\Omega}}_b\T \frac{{\mathbf x} - {\mathbf l}_b}{\|{\mathbf x} - {\mathbf l}_b\|}.
\end{equation}
Throughout this paper, we adopt the unit-vector model introduced in \cite{2016_AbrXiaMarTri} to represent the direction of arrival, as well as the corresponding 3-D directional statistics approach. Errors in estimating the directions of arrival are modeled by using the von Mises-Fisher distribution, which is the correspondent of the 2-D normal distribution to the two-dimensional unit sphere ${\mathcal{S}}^2 \subset {\mathbb{R}}^3$. For a $3 \times 1$ unit vector ${\mathbf u} \in {\mathcal{S}}^2$, the von  Mises-Fisher distribution is given by 
\begin{equation} \label{vmf_distr}
    {\mathrm{VMF}}({\mathbf u} | {\boldsymbol{\mu}}, \kappa)=
    c\:\exp\left(\kappa{\boldsymbol{\mu}}\T{\mathbf u}\right).
\end{equation}
where ${\boldsymbol{\mu}}$ is the mean direction, $\kappa$ is the concentration parameter, and $c=\kappa/(4\pi\sinh\kappa)$ is the normalization constant.
As mentioned earlier, the reason for adopting a directional statistics approach is that the natural parameter space of angles is not an Euclidean space, but a sphere. Angles are periodic in their nature, and therefore, the natural support of the corresponding probability density functions should be the unit sphere.

\section{Probabilistic ToA and AoA Positioning}

\subsection{Probabilistic ToA posiitoning} \label{sec_toa_algorithm}
For ToA localization, the least square (LS) optimization technique is widely used \cite{ToA-NLOS-Tut}. For instance,  the well-known nonlinear LS method solves the optimization below to determine the UD position $\mathbf{x}$ (and the unknown transmit time $\tau$ as a byproduct)
\begin{align}\label{NLS}
(\hat{\mathbf{x}}, \hat{\tau})=&\arg\min_{\mathbf{x},\tau} \sum_{k=1}^K(||\mathbf{p}_{k}-\mathbf{x}||+\tau-t_{k})^2
\end{align}
The main disadvantage of the LS-based approach is that it does not take into account the channel bias errors $\gamma_k$ and thus degrades the localization accuracy in positioning-challenge environments such as urban canyon and indoors.

To overcome the above drawback, in the Bayesian probabilistic ToA localization algorithm \cite{Blade}, the channel bias is incorporated into a MAP estimator as a RV to robustly determine the UD location. Denote the ToA measurement vector for all locators by $\mathbf{t}=[t_1~t_2~...~t_K]^T$. The UD position and the unknown signal transmit time can be estimated as follows (with a non-informative prior $p(\mathbf{x},\tau)$),
\begin{align}\label{MAP}
\hat{\mathbf{x}},\hat{\tau}=\arg\max_{\mathbf{x},\tau}\ln p(\mathbf{t}|\mathbf{x},\tau)
\end{align}
Assuming that the ToA measurements from different locators are independent, the joint log-likelihood is given by  
\begin{align}
\ln p(\mathbf{t}|\mathbf{x},\tau)&=\sum_{k=1}^K\ln p(t_{k}|\mathbf{x},\tau)\nonumber\\
&= \sum_{k=1}^K\ln \int p(t_{k}|\mathbf{x},\tau,\gamma_k)p(\gamma_k)d\gamma_k \label{ll_indpendent}
\end{align}
where $p(t_{k}|\mathbf{x},\tau,\gamma_k)$ is a Gaussian distribution with mean $||\mathbf{x}_{k}-\mathbf{x}||+\tau+\gamma_k$ and variance $\sigma^2$. Given Eq. (\ref{SV-GMM}), the estimator (\ref{MAP}) can be rewritten as  
\begin{align}\label{MAP-SV}
\hat{\mathbf{x}},\hat{\tau}&=\arg\max_{\mathbf{x},\tau}\sum_{k=1}^K\ln p(t_{k}|\mathbf{x},\tau)\nonumber\\
&=\arg\max_{\mathbf{x},\tau}\sum_{k=1}^K\ln\bigg\{\sum_{i=1}^{L_k}\frac{w_{ik}}{\sigma_{ik}\sqrt{2\pi}}\nonumber\\
& \qquad       \exp\left[-\frac{(t_k-||\mathbf{p}_{k}-\mathbf{x}||
-\tau-\mu_{ik})^2}{2\sigma^2_{ik}}\right]\bigg\}
\end{align}
where $\sigma_{ik}^2=\tilde{\sigma}^2_{ik}+\sigma^2$.

\subsection{Probabilistic AoA positioning} \label{sec_aoa_algorithm}

Using the model outlined in Section~\ref{sec_model}, the noisy directional estimates at the $b$-th AoA locator are assumed to have a von mises-Fisher distribution with the mean direction ${\boldsymbol{\mu}}_b=\hat{\mathbf u}_b$, and a concentration parameter $\kappa_b$ whose value reflects the reliability of the estimate.
Given $B$ locators whose directional estimates are $\hat{\mathbf u}_b$, and assuming that they are affected by independent errors, the joint  log-likelihood of the user's position may be expressed using Eqs.~\eqref{true_direction}, and \eqref{vmf_distr}:

\begin{eqnarray} \label{joint_aoa_llf}
    {\mathcal L}_{\angle}({\mathbf x}) 
    &=& \sum_{b=1}^B \ln {\mathrm{VMF}}({\mathbf u}_b({\mathbf x});\hat{\mathbf u}_b,\kappa_b), \nonumber \\
    &=&B\ln c + \sum_{b=1}^B \kappa_b \hat{\mathbf u}_b\T {\boldsymbol{\Omega}}_b\T \frac{{\mathbf x} - {\mathbf l}_b}{\|{\mathbf x} - {\mathbf l}_b\|}
\end{eqnarray}
The user's position can be estimated by maximizing the above joint likelihood, i.e.,
\begin{equation} \label{pos_est_aoa}
\hat{\mathbf x}=\arg\max_{\mathbf x}{\mathcal L}_{\angle}({\mathbf x})
\end{equation}

\subsection{Probabilistic joint positioning}


For the problem of joint ToA and AoA positioning, we assume that the measurements from ToA and AoA locators are all independent. As a result, we could solve the following estimation problem to estimate the user position,
\begin{equation}\label{joint_ML}
    \hat{\mathbf{x}},\hat{\tau} \!=\! \arg\max_{\mathbf{x},\tau} \!
    \left[\sum_{k=1}^K\ln p(t_{k}|\mathbf{x},\tau) \!+\! \sum_{b=1}^B \ln {\mathrm{VMF}}({\mathbf u}_b|\hat{\mathbf u}_b({\mathbf x}),\kappa_b)
    \right]
\end{equation}
Solving the non-convex optimization problem~\eqref{joint_ML} requires a trade-off between convergence speed and computation complexity. Possible solutions include, e.g., gradient ascent~\cite{2016_AbrXiaMarTri}, expectation propagation~\cite{EP}, and Monte Carlo methods~\cite{ToA_AoA_Visa}.

\section{Prototype Setup}\label{sec_system}

\begin{figure*}[tb]
\begin{center}
 \includegraphics[width= 10 cm]{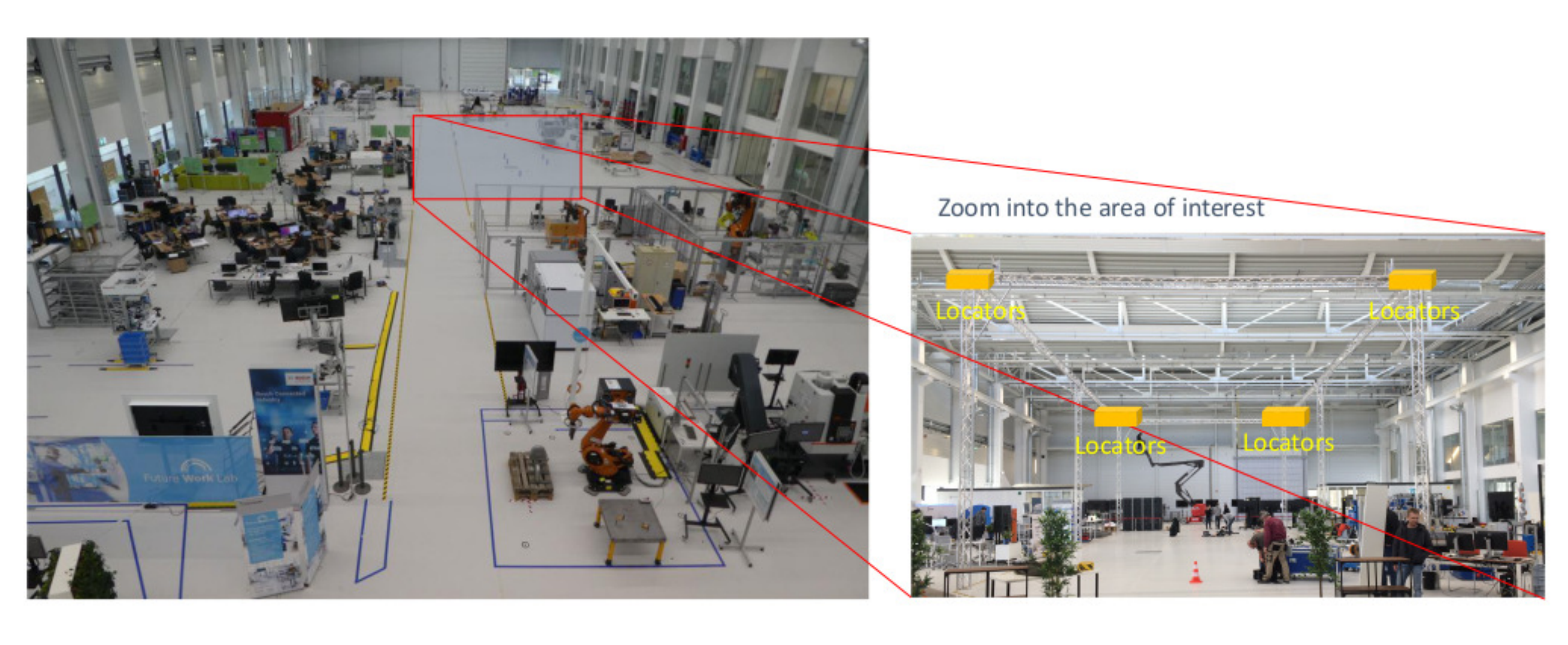}
 \caption{The experiment area in an indoor factory environment. The area of interest is 20 meters $\times$ 10 meters, as shown in the red rectangle in the left figure. At each corner of the area, there is a pair of co-located ToA and AoA locators, as depicted by the yellow cuboids in the right figure. The heights of the locators are around 7.3 meters.}
\label{Arena_pic}
\end{center}
\vspace{-20pt}
\end{figure*}

To evaluate the performance of the probabilistic ToA and AoA joint localization in real indoor environments, we build a prototype based on proprietary ToA and AoA localization systems in the ARENA2036 research building,\footnote{For the proprietary ToA localization system, a similar hardware setup is used in 5GCAR project for outdoor positioning \cite{OSF20}.}  which aims to offer a realistic indoor factory environment for developing and testing concepts of future transport \cite{Arena2036}. The entire ARENA2036 building is 130m long and 46m wide, with a sawtooth roof and folded aluminum fa\c{c}ade (as shown in Fig.~\ref{Arena_pic}). In our experiment, we focus on a smaller area inside with a size of 20m$\times$10m. In each corner of this area, a pair of co-located ToA and AoA locators was installed with height around 7.3m. 




In our prototype, the UD broadcasts wideband Pseudo-Noise (PN) sequences (with length 4096) and BlueTooth Low Energy (BLE) signals as the ToA and AoA positioning reference signals, respectively. The ToA reference signal bandwidth is around 50 MHz. At each ToA locator, only one antenna is employed to receive signals.  Sliding correlators with thresholding \cite{Blade_WCNC} are used to measure the ToA of reference signals for each ToA locator. To achieve high synchronization accuracy, all 4 ToA locators are synchronized with a central server using the White Rabbit (WR) protocol \cite{WR}. In our prototype, the synchronization error is less than 1 nanosecond. The lengths of the cables connecting different ToA locators to the central server are carefully measured and subtracted from the measured ToAs to derive the final ToA values used for positioning. At the AoA locators, the AoA estimation is obtained by using a planar antenna array with 7 dual-polarized elements that receives BLE signals. 




\section{Experiment Results}\label{sec_exp}

\begin{figure}[t]
\begin{center}
 \includegraphics[width=7 cm ]{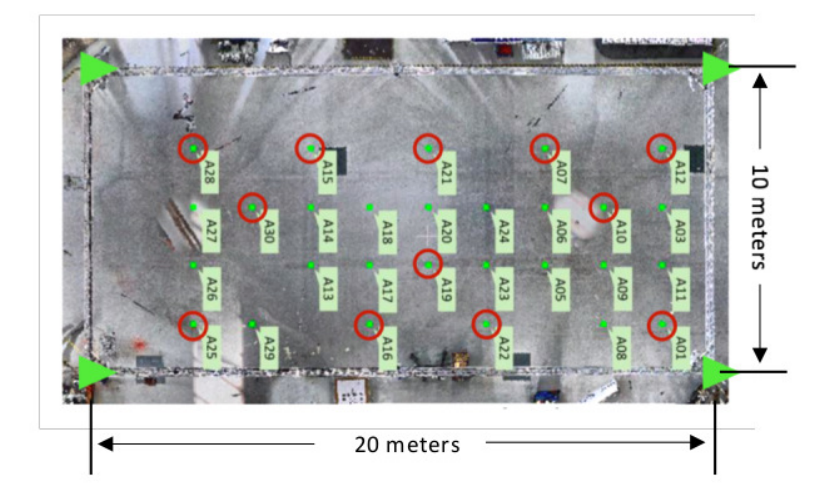}
 \caption{The top-down view of the experiment area. The green triangles and dots indicate the positions of the locators and TPs, respectively.}
\label{Arena_top}
\end{center}
\vspace{-20pt}
\end{figure}

In this section, we use over-the-air measurements from the proprietary joint ToA and AoA localization system in the ARENA2036 experiment area (described in Section \ref{sec_system}) to evaluate the positioning performance in real indoor environments.  As shown in Fig. \ref{Arena_top}, there are totally 28 test points (TPs) in that area. We took both ToA and AoA measurements at each TP. To avoid sophisticated training of the probabilistic models, in this study we heuristically choose the following parameters for the probabilistic localization algorithms: $L_k=1$, $\mu_{1k}=0$, $\tilde{\sigma}_{1k}^2=1$, $\sigma^2=10^{-5}$, and $\kappa=10$, where $k\in\{1,2,3,4\}$.\footnote{It is possible to improve the localization performance by learning and fine-tuning the parameters in the probabilistic models, which, however, is out the scope of this paper. For instance, see \cite{Blade, ToA-MVGMM} for training GMM in ToA localization.}



\begin{figure}[tb]
\begin{center}
 \includegraphics[width= 6.1 cm]{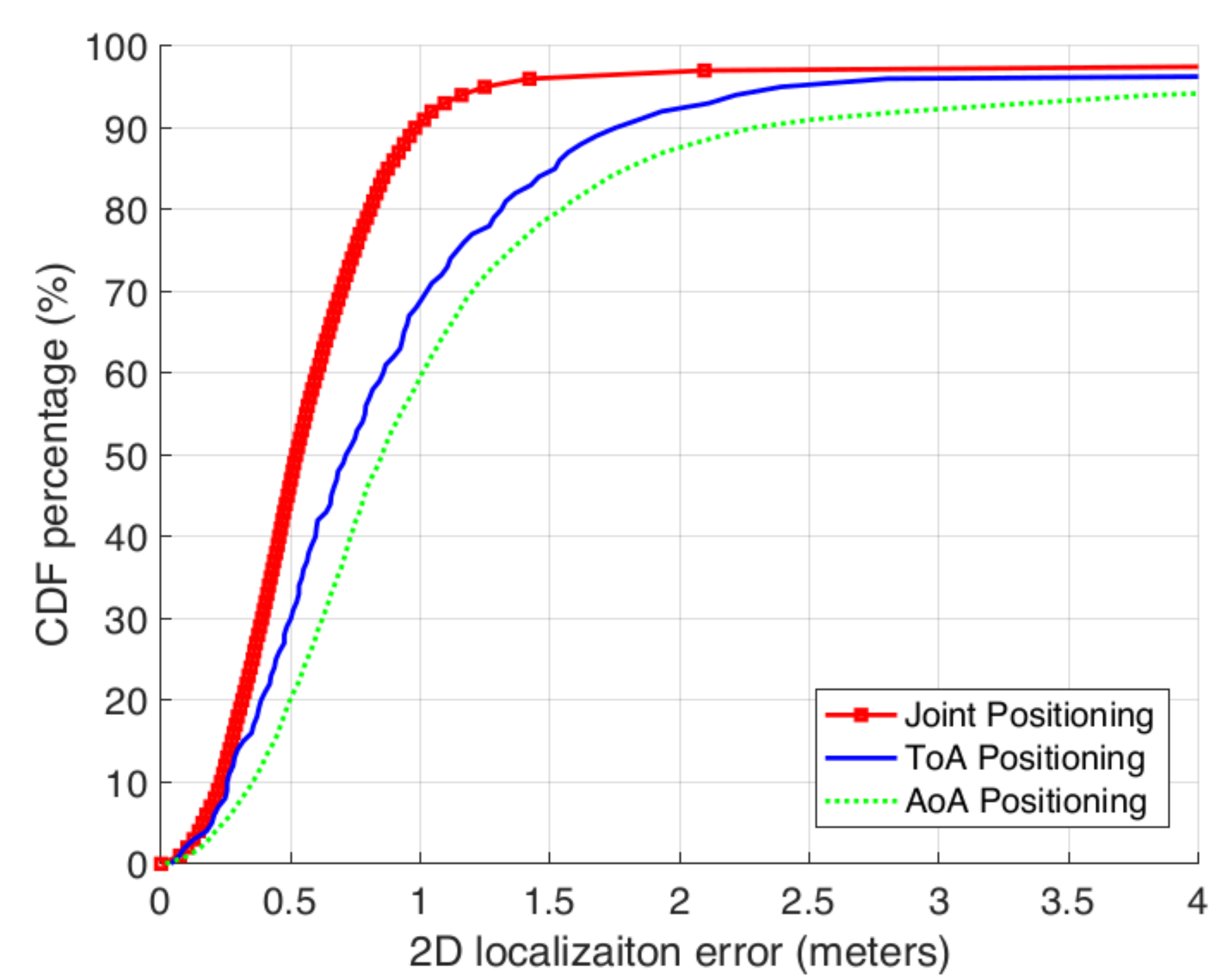}
 \caption{The CDF of horizontal localization errors for different localization algorithms in the field test. Note that the curves correspond to unfiltered (raw) measurements.} 
\label{err_cdf}
\end{center}
\vspace{-20pt}
\end{figure}

\begin{figure*}[tb]
\begin{center}
 \includegraphics[width= 9 cm]{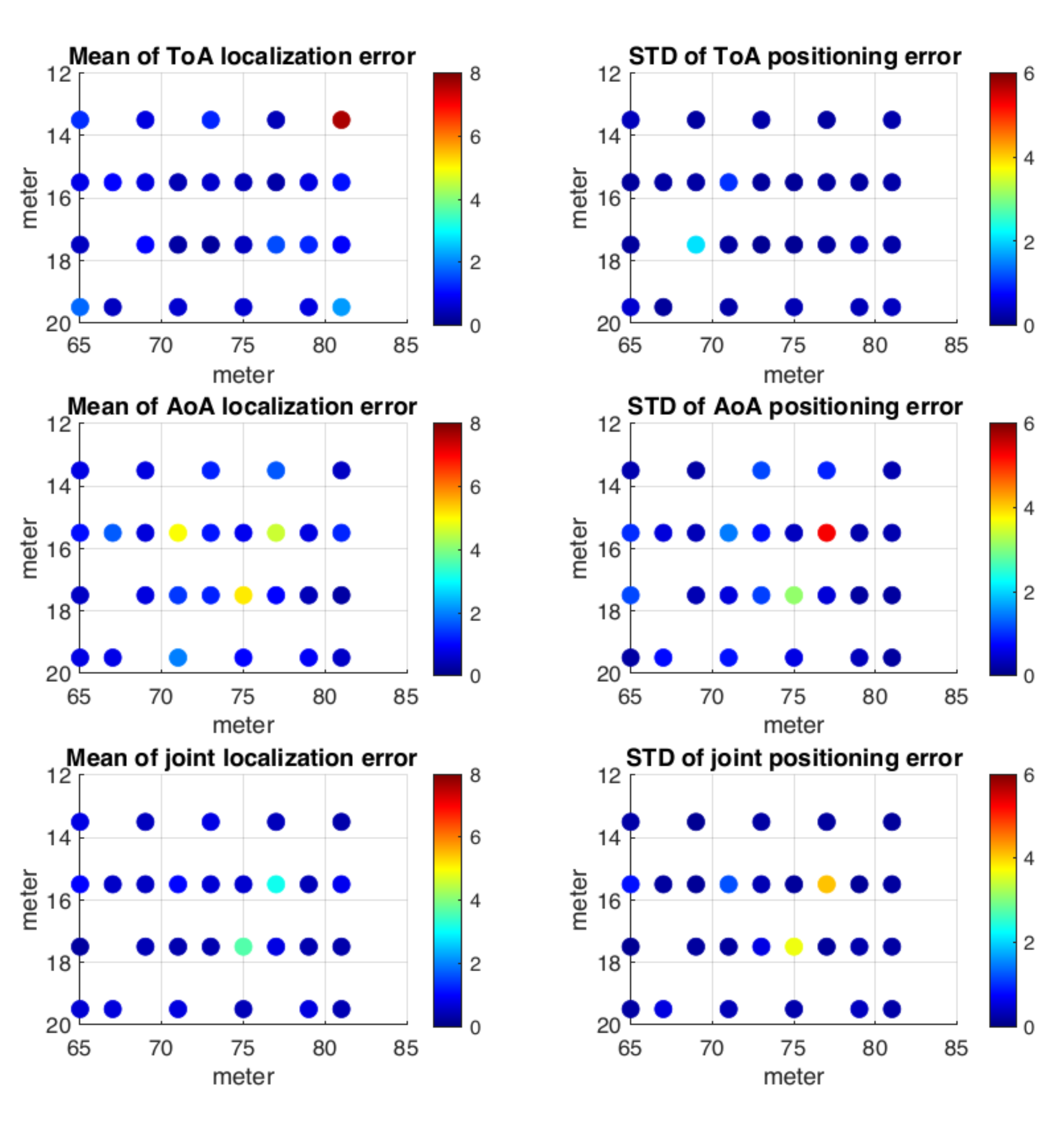}
 \caption{Mean and standard deviation (STD) of horizontal localization errors at each TP for different positioning algorithms: TOA only -- top row, AoA only -- middle row, joint ToA+AoA -- bottom row. Colorbar encodes the mean and the standard deviation values in meters.}
\label{heatmap}
\end{center}
\vspace{-20pt}
\end{figure*}

We compare the localization performance of the joint positioning algorithm with the baseline approaches using either ToA or AoA data only. Specifically, for the baseline with AoA measurements only, we use the probabilistic algorithm presented in Section \ref{sec_aoa_algorithm}. For the baseline with ToA measurements only, we note that in our filed test since the ToA probabilistic model only includes a single Gaussian component, the nonlinear LS approach achieves similar or slightly better performance compared with the probabilistic algorithm in Section \ref{sec_toa_algorithm} with parameters mentioned before in this section. As a result, we adopt the nonlinear LS algorithm as the ToA baseline here. Fig.~\ref{err_cdf} depicts the empirical cumulative distribution functions (CDF) of horizontal localization error for all three algorithms in the field test. It shows that the joint positioning algorithms significantly outperforms the baselines. A  detailed comparison is given in Table \ref{tab:sim_com}. As shown, in terms of mean, RMS, $50\%$-ile and $90\%$-ile localization errors, the joint positioning algorithm outperforms the ToA baseline by $29.3\%$, $15.3\%$, $26.9\%$, and $44.0\%$, respectively, and the AoA baseline by $43.8\%$, $32.9\%$, $38.8\%$, and $57.0\%$, respectively. The overall localization performance depends on key factors such as: the density and geometry of the deployed locators, antenna array, time synchronization, and propagation environment. 

\begin{center}
\captionof{table}{Horizontal Localization Error} \label{tab:sim_com} 
  \begin{tabular}{  p{1.5cm} | p{1.7cm} | p{1.7cm}  |p{1.7cm} }
    \hline
     & Joint   & ToA-only  & AoA-only    \\ \hline
    Mean & 0.763m & 1.079m & 1.357m \\ \hline
    RMS & 1.528m & 1.803m & 2.276m\\  \hline
    50\% CDF & 0.522m & 0.714m & 0.853m\\ \hline
    90\% CDF & 0.981m & 1.753m & 2.282m\\  \hline
  \end{tabular}
\end{center}
\vspace{10pt}



Fig.~\ref{heatmap} further depicts the mean and standard deviation (STD) of horizontal localization errors for all three positioning algorithms at each TP. One can find that for the joint positioning algorithm, in almost all TPs, the mean and STD of localization errors are both in the sub-meter level (not the case for ToA-only and AoA-only baselines). The long tail of the joint positioning error in Fig.~\ref{err_cdf} mainly comes from A06 and A23, which is due to the severe NLOS and multipath errors which are not well-captured in the probabilisitic model.\footnote{We expect that a more sophisticated well-trained model could further improve the localization performance.} 

\begin{figure}[tb]
\begin{center}
 \includegraphics[width= 6 cm]{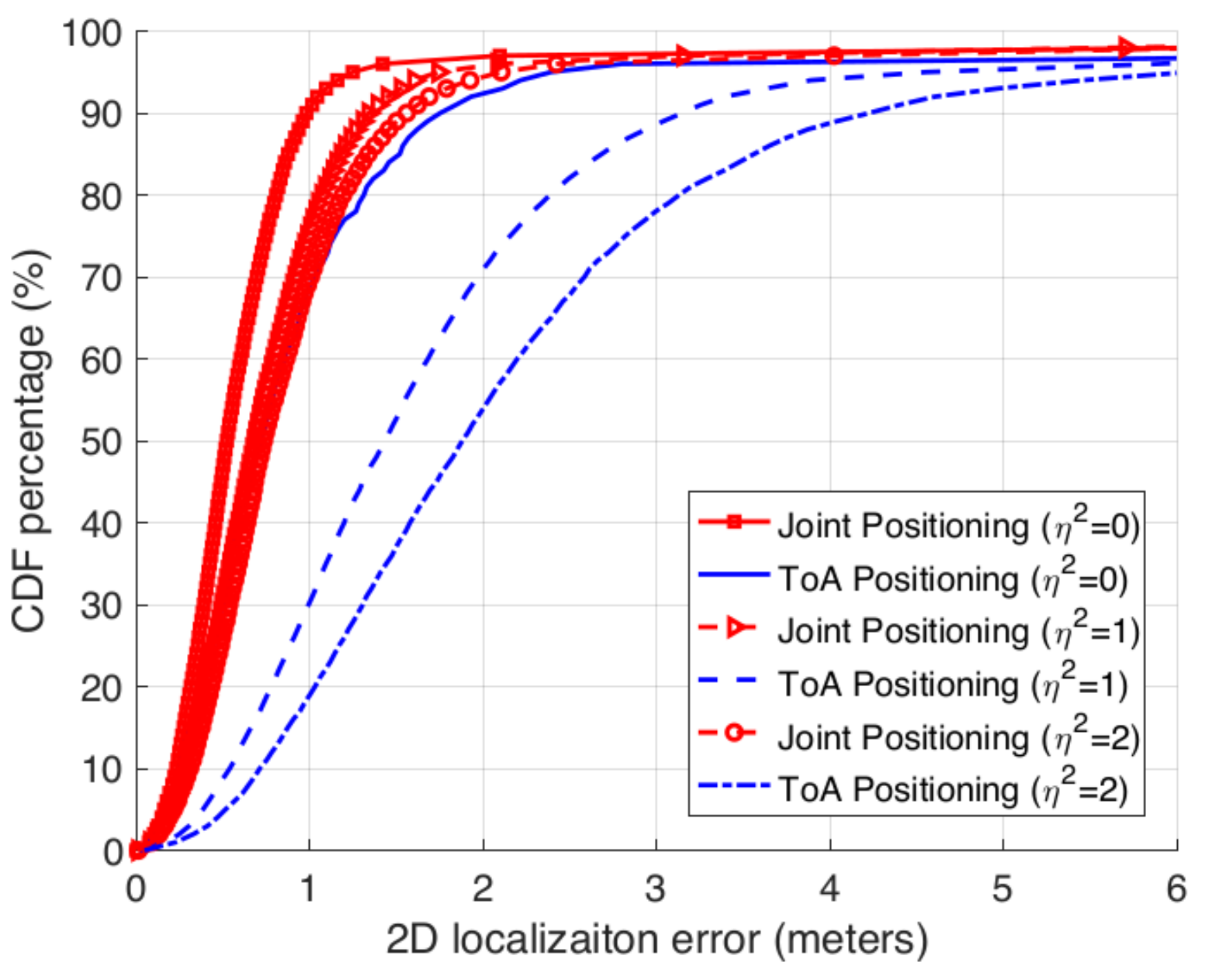}
 \caption{The impact of synchronization errors on joint ToA+AoA positioning and the ToA baseline, where $\eta$ is the standard deviation of the zero-mean Gaussian synchronization error, with the unit meter (i.e., converted from time by multiplying with the speed of light).}
\label{timing}
\end{center}
\vspace{-20pt}
\end{figure}

It is well known that synchronization among locators is essential for ToA localization. In practice, it is challenging to maintain accurate synchronization, unless some sophisticated synchronization protocols (e.g., WR in our prototype) are applied. One may wonder how larger synchronization errors affect the performance of joint ToA and AoA localization. To model additional synchronization errors (besides the errors already induced by the WR protocol in real measurements), we add each ToA measurement together with an i.i.d Gaussian RV of mean 0 and variance $\eta^2$. Fig.~\ref{timing} shows the CDF of horizontal localization error for the joint localization algorithm and the ToA baseline, with respect to different $\eta^2$ values. One can find that by leveraging additional AoA data, the former is much more robust to synchronization errors than the latter.

\section{Conclusion} 
In this work, we advocate a joint probabilistic ToA and AoA 3-D localization algorithm and evaluate its performance in an indoor factory environment with proprietary localization systems.  The prototype is with 4 pairs of co-located ToA and AoA locators, each of which is at a corner at the ceiling of a 10 meter by 20 meter section, where the WR protocol  is employed to achieve high accuracy synchronization among ToA locators. In the joint localization algorithm, only ToA and AoA measurements are needed, and the low-level information like CSI is not required. To mitigate multipath, the ToA channel bias error and angular uncertainty (due to multipath and NLOS reflections) are modeled as RVs following GMM and von Mises-Fisher distributions, respectively. In the field test, the joint localization algorithm is able to achieve sub-meter level accuracy  at $90\%$-ile for horizontal localization, which significantly outperforms the baselines using either ToA or AoA data individually. In addition, we numerically assess how different synchronization errors affect the performance of the joint localization algorithm. It turns out that, compared with the baseline using only ToA measurements, the joint localization algorithm is much more robust to synchronization errors by leveraging additional AoA data. Future work includes evaluating the algorithm comprehensively in more indoor scenarios, and generalizing the algorithm to reject outlying measurements and track mobile users.

\section*{Acknowledgment}

The authors would like to thank the colleagues Thomas Ahlich, Silvio Mandelli, Stephan Saur, Thomas Schlitter and Maik Steudtner for taking the measurements in ARENA2036 and technical discussions.



%



\bibliographystyle{IEEEtran}
\bibliography{loc}


\end{document}